\documentclass[10pt,journal,comsoc]{IEEEtran}

\usepackage{cite}
\usepackage{amsmath,amssymb,amsfonts}
\usepackage[normalem]{ulem}
\usepackage{algorithmicx}
\usepackage{graphicx}
\usepackage{textcomp}
\usepackage{svg}      
\usepackage{xcolor}
\usepackage[most]{tcolorbox}
\usepackage[table]{xcolor}
\usepackage[hidelinks]{hyperref}
\usepackage{listings}
\usepackage{multirow}
\usepackage{graphicx}     
\usepackage{caption}      
\tcbset{colback=gray!10!white, colframe=gray!30!black, boxrule=0.4pt, arc=2pt,
        left=6pt,right=6pt,top=6pt,bottom=6pt, fonttitle=\bfseries}
\usepackage{subcaption}   
\usepackage{comment}
\usepackage{pifont}
\usepackage{booktabs}
\usepackage{colortbl}
\usepackage{array}
\usepackage{float}

\tcbset{
  colback=white, colframe=black!35,
  boxsep=1pt, left=4pt, right=4pt, top=3pt, bottom=3pt,
  before skip=4pt, after skip=4pt,
  fonttitle=\bfseries\footnotesize
}
\usepackage[linesnumbered, ruled]{algorithm2e}

\SetKwRepeat{Do}{do}{while}%

\SetCommentSty{mycommfont}

\usepackage{algpseudocode}

\makeatletter
\newcommand{\algmargin}{\the\ALG@thistlm}
\makeatother

\newlength{\forwidth}
\algdef{SE}[parFOR]{parFor}{EndparFor}[1]
{\parbox[t]{\dimexpr\linewidth-\algmargin}{%
\hangindent\forwidth\strut\algorithmicfor\ #1\ \algorithmicdo\strut}}{\algorithmicend\ \algorithmicfor}%

\newlength{\whilewidth}
\algdef{SE}[parWHILE]{parWhile}{EndparWhile}[1]
{\parbox[t]{\dimexpr\linewidth-\algmargin}{%
\hangindent\whilewidth\strut\algorithmicwhile\ #1\ \algorithmicdo\strut}}{\algorithmicend\ \algorithmicwhile}%
\algnewcommand{\parState}[1]{ %
\parbox[t]{\dimexpr\linewidth-\algmargin}{\strut #1\strut}}

\usepackage{epstopdf}

\usepackage{titlesec}
\titlespacing*{\section}{0pt}{0.7\baselineskip}{0.4\baselineskip}
\titlespacing*{\subsection}{0pt}{0.5\baselineskip}{0.3\baselineskip}
\titlespacing*{\subsubsection}{0pt}{0.4\baselineskip}{0.25\baselineskip}
\begin{document}

\title{FTA-NTN: Fairness and Throughput Assurance in Non-Terrestrial Networks}
\author{
\IEEEauthorblockN{Sachin Ravikant Trankatwar, Heiko Straulino, Petar Djukic, Burak Kantarci}\\
\thanks{Sachin Trankatwar and Burak Kantarci are with University of Ottawa, Ottawa, ON, Canada. Emails: \{sravikan,burak.kantarci\}@uottawa.ca\\
 Heiko Straulino and Petar Djukic are with Nokia Bell Labs, 600 March Road,
 Kanata, ON K2K 2E6, Canada. Email: \{heiko.straulino, petar.djukic\}@nokia-bell-labs.com 
}
}

\markboth{}%
{}

\IEEEtitleabstractindextext{

\begin{abstract}
Designing optimal non-terrestrial network (NTN) constellations is essential for maximizing throughput and ensuring fair resource distribution. 
This paper presents FTA-NTN (Fairness and Throughput Assurance in Non-Terrestrial Networks), a multi-objective optimization framework that jointly maximizes throughput and fairness under realistic system constraints. 
The framework integrates multi-layer Walker Delta constellations, a parametric mobility model for user distributions across Canadian land regions, adaptive K-Means clustering for beamforming and user association, and Bayesian optimization for parameter tuning. 
Simulation results with $500$ users show that FTA-NTN achieves over $9.88~\text{Gbps}$ of aggregate throughput with an average fairness of $0.42$, corresponding to an optimal configuration of $9$ planes with $15$ satellites per plane in LEO and $7$ planes with $3$ satellites per plane in MEO. 
These values align with 3GPP NTN evaluation scenarios and representative system assumptions, confirming their relevance for realistic deployments. Overall, FTA-NTN demonstrates that throughput and fairness can be jointly optimized under practical constraints, advancing beyond throughput-centric designs in the literature and offering a scalable methodology for next-generation NTN deployments that supports efficient and equitable global connectivity.
\end{abstract}

\begin{IEEEkeywords}
Non-Terrestrial Networks (NTN), Satellite constellations (LEO/MEO/GEO), Multi-objective optimization (MOO), Beamforming and resource allocation, Fairness--throughput trade-off
\end{IEEEkeywords}}
\maketitle

\IEEEdisplaynontitleabstractindextext
\IEEEpeerreviewmaketitle
\thispagestyle{empty}
\pagestyle{empty}
\section{Introduction}\label{Intro}
Non-terrestrial networks (NTNs) have emerged as a promising solution to extend connectivity to remote, underserved, and mobile environments where terrestrial infrastructure is limited or unavailable~\cite{9210567,9275613,nguyen2024emerging}. With the growing demand for global broadband access, the design and operation of satellite-based communication systems have become increasingly critical~\cite{9210567,9275613,nguyen2024emerging}. A key challenge in NTN deployment lies in jointly maximizing system throughput and ensuring fairness in resource distribution, given the dynamic nature of user demand and satellite coverage~\cite{10279350,10551694}. Achieving this balance requires careful constellation design, particularly with respect to the number of orbital planes and satellites per plane~\cite{rs12111845,ko2025satellite}, while accounting for practical limitations such as beam capacity, user association constraints, and the need for scalable beam management~\cite{chen2024beam,zhang2025framework}.

\noindent
Despite extensive NTN research, jointly maximizing throughput and fairness under realistic constraints remains an open challenge, particularly when constellation design must account for limited beam capacity, user--beam association restrictions, and dynamic user distributions. 
To address this problem, we propose Fairness and Throughput Assurance in Non-Terrestrial Networks (FTA-NTN), a multi-objective optimization (MOO) framework tailored to satellite constellations.  
The main contributions of this work are summarized as follows:
\begin{itemize}
    \item A scalable optimization framework integrating multi-layer \textit{Walker-Delta} constellations, a parametric Canadian mobility model, and adaptive K-Means clustering for beam assignment and user association.
    \item A joint throughput--fairness objective reformulated via a weighted-sum approach and solved efficiently using \textit{Bayesian optimization} under user- and beam-level constraints.
\end{itemize}
\noindent
Extensive simulations with dynamic user scenarios are conducted to evaluate the proposed framework. 
Results show that FTA-NTN achieves over $9.88~\text{Gbps}$ of aggregate throughput with an average fairness index of $0.42$ using $500$ users, converging to optimal LEO and MEO configurations. 
These values align with 3GPP NTN evaluation scenarios and representative link-budget assumptions~\cite{3gpp38811,3gpp38821}, confirming their relevance for realistic deployments. Together, these outcomes validate FTA-NTN as a practical design methodology that advances beyond throughput-centric approaches in the literature, enabling scalable and equitable next-generation NTN deployments.

\noindent
The rest of the paper is organized as follows. Section~\ref{RelWork} presents a comprehensive review of related works. In Section~\ref{SysMod}, we describe the system model and formulate the MOO problem. Section~\ref{Sol} outlines our proposed solution. Simulation results and performance evaluations of the proposed scheme are discussed in Section~\ref{Res}. Finally, Section~\ref{Conc} concludes the paper with key insights and directions for future research.

\section{Related Work}\label{RelWork}
Recent publications collectively underscore the dual role of NTNs as both a technological frontier and a key enabler for 6G. In paper~\cite{9210567}, examine integration and standardization aspects, ~\cite{9275613} highlights strategic 6G challenges such as mobility and resource management, and work~\cite{nguyen2024emerging} emphasizes emerging enablers like AI, spectrum, and slicing. Together, they frame NTNs as central to future global connectivity.

\noindent
The study in~\cite{10279350} proposes a beam-level resource allocation framework for LEO constellations, while~\cite{10551694} addresses fairness-aware throughput optimization in LEO satellite networks using a 3D spherical coordinate system for user association. Optimal LEO-based global navigation and augmentation constellations are designed in~\cite{rs12111845} using Genetic Algorithms and NSGA-II to improve coverage and positioning accuracy. An integer linear programming approach for constellation design is introduced in~\cite{ko2025satellite} to minimize worst-case revisit time. The work in~\cite{chen2024beam} presents an adaptive clustering strategy with large, medium, and small footprints to reduce latency and enhance throughput. A joint beam scheduling and resource allocation framework based on beam-hopping is explored in~\cite{zhang2025framework}, enabling multi-domain optimization across spatial, temporal, frequency, and power resources. Network performance of high-inclination Walker Delta pLEO constellations is studied in~\cite{10115752}, focusing on path failure rate, latency, and routing under different ISL topologies, while~\cite{jeon2024communication} investigates continuous coverage and ISL connectivity. Beam management and handover control in LEO systems are addressed in~\cite{zhu2024beam}, emphasizing minimizing user disruption and supporting spectrum coexistence with terrestrial networks. In~\cite{9149179} provides a detailed link-budget analysis for NTNs, focusing on system-level performance and the impact of satellite channel impairments on PHY/MAC procedures. Finally,~\cite{9439912} employs reinforcement learning-based methods for constellation-level multi-objective resource allocation.

\noindent
While recent publications~\cite{9210567,9275613,nguyen2024emerging} provide valuable insights into the evolving New Space ecosystem and outline broad research directions, they do not present concrete optimization frameworks. Prior works~\cite{10279350,10551694,rs12111845,ko2025satellite,chen2024beam,zhang2025framework,jeon2024communication,zhu2024beam,9439912,10115752,9149179} largely address individual performance aspects---such as beam management or user association---without jointly optimizing throughput and fairness. Moreover, these studies often assume fixed constellation configurations and primarily consider LEO satellites, overlooking the synergistic roles of MEO and GEO layers in a fully integrated NTN. To bridge these gaps, we introduce FTA-NTN, a comprehensive MOO framework that jointly maximizes throughput and fairness while dynamically optimizing key constellation parameters. Unlike prior approaches, FTA-NTN incorporates realistic system constraints and leverages the combined benefits of LEO, MEO, and GEO, offering a scalable and deployment-ready design methodology for future NTN architectures.

\section{System Model and Problem Formulation}\label{SysMod}
This section introduces the FTA-NTN system architecture and formulates a MOO problem that jointly maximizes throughput and fairness. 
\subsection{System Model}\label{sec:system_model}
\begin{figure}[htbp]
    \centering
    \includegraphics[width=0.75\linewidth]{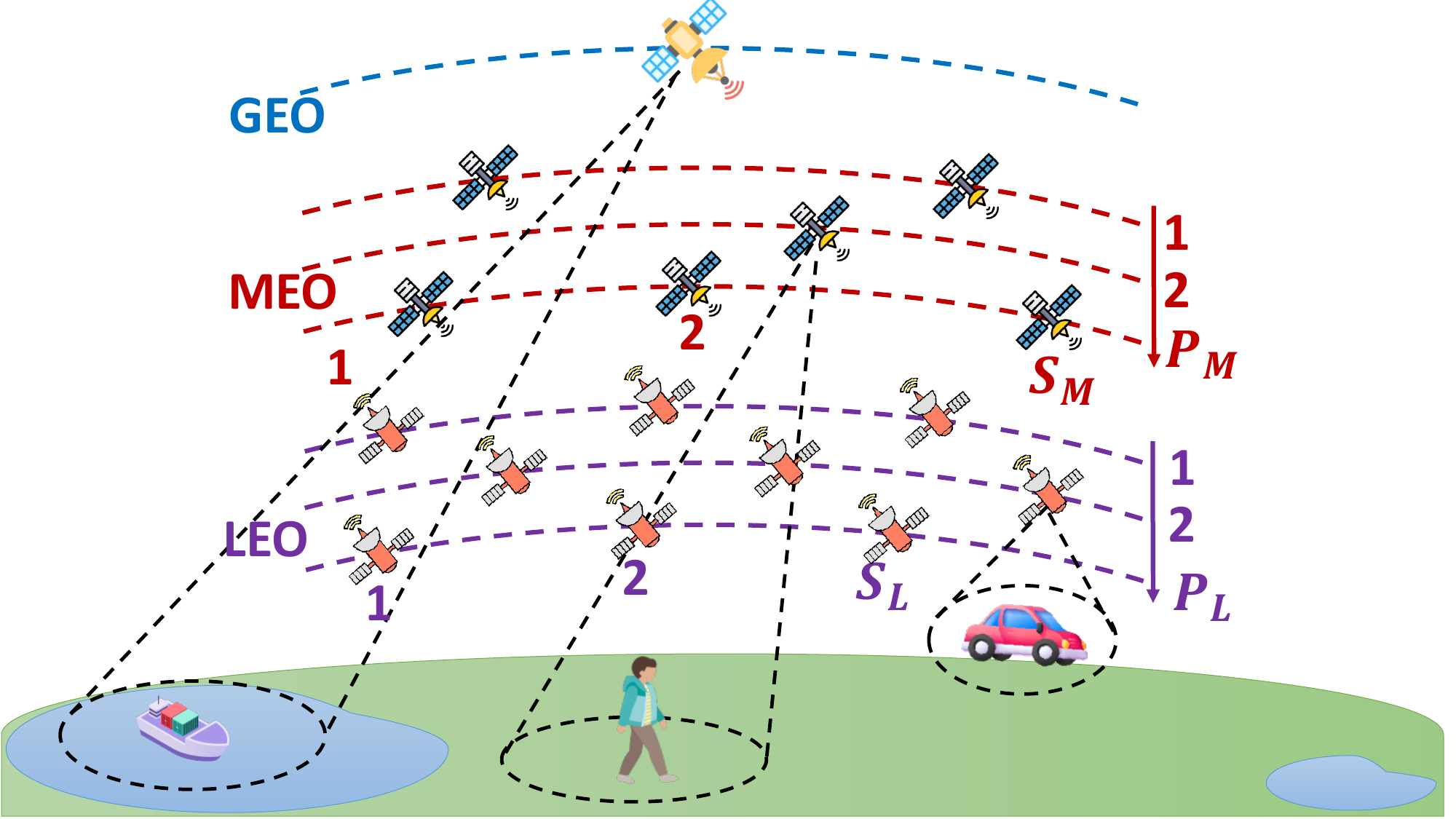}
    \caption{The NTN system  with three orbital layers.}
    \label{fig:system_model}
\end{figure}
We consider a downlink FTA-NTN multi-layer comprising LEO, MEO, and GEO, denoted by $k \in \{L, M, G\}$, as illustrated in Fig.~\ref{fig:system_model}. Each layer $k$ is defined by $(P_k, S_k, h_k)$, representing the number of planes, satellites per plane, and orbital altitude; for example, $(P_L, S_L)$ and $(P_M, S_M)$ for LEO and MEO. With fewer satellites due to its altitude and larger coverage, the GEO layer is modeled separately without orbital planes. $\mathcal{S}$ denotes the full satellite set.  
Each satellite $s \in \mathcal{S}$ supports up to $X$ spot beams, each serving at most $Z$ users. The set of beams for $s$, denoted $\mathcal{B}_s$, is dynamically positioned through user clustering, and a beam is activated only if the satellite has not exceeded its beam limit $X$. The network serves $N$ mobile users, $\mathcal{U} = \{1, 2, \dots, N\}$, distributed across Canadian land regions and moving over time.  
Users are associated with one of the three orbital layers at each time step based on coverage availability. We denote by $\mathcal{U}_k \subseteq \mathcal{U}$ the subset of users served by layer $k$. Beam--user associations are represented by a binary matrix $A = \{A_{ij}\}$, where $A_{ij} = 1$ if user $i$ is served by beam $j$, and $A_{ij} = 0$ otherwise.  
The system is modeled through three components described in the following sections, forming the basis for the problem formulation.
\subsubsection{Satellite Constellation Modeling}
Each orbital layer $k$ is modeled using a Walker-delta constellation, parameterized by $(P_k, S_k, h_k)$. The semi-major axis of layer $k$ is $a_k = R_{\oplus} + h_k$, where $R_{\oplus} = 6371$ km is Earth’s mean radius. To ensure uniform constellation geometry, the right ascension of the ascending node (RAAN) of plane $p$ and the mean anomaly of satellite $s$ in that plane are defined as
\begin{equation}
    \Omega_{k,p} = \frac{2\pi p}{P_k}, 
    \qquad
    M_{k,p,s} = \left(\frac{2\pi s}{S_k} + \frac{\pi p}{S_k}\right) \bmod 2\pi,
\end{equation}
with $p = 0,1,\dots,P_k-1$ and $s = 0,1,\dots,S_k-1$. Here, $\Omega_{k,p}$ governs the angular separation of orbital planes, while $M_{k,p,s}$ ensures uniform phasing and along-track separation of satellites within and across planes.

\noindent
Satellites are initialized at a reference epoch $t_0$ and propagated over time using two-body orbital dynamics. At each time step $t$, the Cartesian states are transformed into geodetic coordinates, yielding geodetic latitude $\phi_{k,p,s}(t)$, geodetic longitude $\lambda_{k,p,s}(t)$, and $h_k$. 
The overall satellite set is expressed as
\begin{equation}
    \mathcal{S} = \bigcup_{k \in \{L,M,G\}} 
    \bigcup_{p=0}^{P_k-1} \bigcup_{s=0}^{S_k-1} 
    \{ (k,p,s), \phi_{k,p,s}(t), \lambda_{k,p,s}(t), h_k \},
\end{equation}
where each element specifies the layer, plane, and satellite index together with its instantaneous geodetic position. This structure reflects the configurations commonly adopted in real-world systems such as Starlink, OneWeb, and Galileo, thereby ensuring both realism and practical relevance.

\subsubsection{Parametric User Mobility Modeling}
To capture end-user dynamics within FTA-NTN, we employ a mobility model inspired by Spatio-Temporal Parametric Stepping (STEPS)~\cite{nguyen2011steps}, distributing $N$ users uniformly across the Canadian landmass with land-only placement. Each user is initialized with a velocity vector $\mathbf{v}_i$ drawn from a uniform distribution and an acceleration vector $\mathbf{a}_i$ from a Gaussian distribution. User trajectories evolve steps $t$ via the kinematic update:
\begin{equation}
    \mathbf{v}_i(t+1) = \mathbf{v}_i(t) + \mathbf{a}_i + \boldsymbol{\eta}, 
    \qquad 
    \mathbf{x}_i(t+1) = \mathbf{x}_i(t) + \mathbf{v}_i(t+1),
\end{equation}
where $\boldsymbol{\eta}$ represents stochastic perturbations and 
$\mathbf{x}_i(t) = \big(\phi_i(t), \lambda_i(t)\big)$ denotes the geodetic latitude–longitude position of user $i$. The resulting traces generate latitude–longitude paths for all users, producing a spatio-temporal demand distribution $\mathcal{U}$ that captures wide-area mobility across Canada and provides a flexible basis for NTN performance evaluation under land-restricted, dynamically evolving demand.

\subsubsection{Path Loss and Interference Modeling}\label{PL}
We model NTN signal degradation through \textit{Free Space Path Loss (FSPL)} and co-frequency \textit{interference}. FSPL quantifies line-of-sight attenuation and is given by, 
\begin{equation}\label{eqn:FSPL}
\text{FSPL}{\text{(dB)}} = 20 \log_{10}(d_{\text{km}}) + 20 \log_{10}(f_{\text{GHz}}) + 92.45
\end{equation}
where $d_{\text{km}}$ is the slant range between satellite and user in kilometers (km), and $f_{\text{GHz}}$ is the carrier frequency. To compute $d_{\text{km}}$, we use the Haversine formula~\cite{sinnott1984virtues}:
{\footnotesize
\begin{equation}\label{eqn:Haversine}
d_{\text{km}} = 2R_{\oplus} \cdot \arcsin \left( \sqrt{ \sin^2\left( \frac{\Delta \phi}{2} \right) + \cos(\phi_1) \cos(\phi_2) \sin^2\left( \frac{\Delta \lambda}{2} \right) } \right)
\end{equation}
}

\noindent
where $\Delta \phi$ and $\Delta \lambda$ are the latitude and longitude differences in radians, and $(\phi_1, \lambda_1)$, $(\phi_2, \lambda_2)$ are the coordinates of the user and satellite, respectively.
Excess Path Loss (EPL) accounts for additional signal attenuation caused by atmospheric impairments such as gas absorption, rain fading, and shadowed multipath propagation. To account for atmospheric attenuation, we include the EPL in dB as $\text{EPL}{\text{(dB)}} = L_{\text{rain}} + L_{\text{cloud}} + L_{\text{vapor}}.$ 
The path loss (PL) gives the complete signal attenuation experienced by the user, 
{\begin{equation}\label{eqn:PL}
\text{PL}_{\text{Tot}} = \text{FSPL}{\text{(dB)}} + \text{EPL}{\text{(dB)}} 
\end{equation}}

\noindent
Accordingly, the received signal power at the user terminal is,
{\begin{equation}\label{eqn:ReceivedPr}
P_r^{\text{dBm}} = P_t + G_t + G_r - \text{PL}_{\text{Tot}}
\end{equation}}
where $P_t$ is the transmit power, $G_t$ is satellite antenna gain, and $G_r$ is user received gain.
Noise power is modeled as, $N_o = 10 \log_{10}(k T B \cdot 10^3) + \text{NF} \quad [\text{dBm}],$
where $k$ is Boltzmann's constant, $T = 290~\text{K}$, $B$ channel bandwidth, and $\text{NF}$ is noise figure.
The received interference power from any interfering satellite beam is given by, 
{\begin{equation}\label{eqn:ReceivedIntPr}
P_{\text{INT}}^{\text{dBm}} = P_t + G_t^{\text{eff}} + G_r - \text{FSPL}_{\text{INT}}
\end{equation}}
where $G_t^{\text{eff}}$ is the effective transmit gain of the interfering beam and $\text{FSPL}_{\text{INT}}$ is computed using the slant distance {(Eq.~\ref{eqn:Haversine})}; from the interfering satellite to the user.
Interference is classified as intra-satellite $(I_{\text{intra},i,k})$ and inter-satellite $(I_{\text{inter},i,k})$~\cite{3gppTR38821}. For a user $i$ in layer $k$,
\begin{equation}\label{eqn:IntraInterf}
I_{\text{intra},i,k} = \sum_{j \in \mathcal{S}_{\text{same},k}} 10^{P^{\text{dBm}}_{\text{INT},j}/10},
\end{equation}
\begin{equation}\label{eqn:InterInterf}
I_{\text{inter},i,k} = \sum_{j \in \mathcal{S}_{\text{other},k}} 10^{P^{\text{dBm}}_{\text{INT},j}/10},
\end{equation}
where $\mathcal{S}_{\text{same},k}$ is the set of beams on the serving satellite, $\mathcal{S}_{\text{other},k}$ the beams from all other visible satellites, and $P_{\text{INT},j}^{\text{dBm}}$ is the received interference power in dBm from the $j^\text{th}$ beam. Thus, the total interference power for user \(i\) in layer \(k\) is $I_{\text{total},i,k} = I_{\text{intra},i,k} + I_{\text{inter},i,k}$.  
Finally, the SINR for user $i$ in layer $k$ is
\begin{equation}\label{eqn:SINR}
\text{SINR}_{i,k} = \frac{10^{P_r^{\text{dBm}}/10}}{I_{\text{total},i,k} + 10^{N_o/10}}.
\end{equation}

\subsubsection{Sum Rate in FTA-NTN}
The system-wide sum rate is evaluated by aggregating the throughput contributions from all users served across each satellite layer \(k\). This formulation captures the combined data delivery capacity of all satellite layers in the integrated NTN architecture.
Let the set of layers be \(\mathcal{K}=\{L,M,G\}\) and \(B_{i,\ell}\) the bandwidth allocated to user \(i\) in layer \(\ell\). The total system-wide sum rate is compactly expressed as~\cite{trankatwar2024power}:
\begin{equation}\label{eq:Rtot}
R_{\text{total}} \;=\; \sum_{\ell \in \mathcal{K}} \;\sum_{i \in \mathcal{U}_\ell}
B_{i,\ell}\,\log_{2}\!\big(1+\mathrm{SINR}_{i,\ell}\big).
\end{equation}

\subsubsection{Fairness in FTA-NTN}
To evaluate user-level fairness in the NTN, we adopt Jain's Fairness Index (JFI), a widely used metric that quantifies the uniformity of throughput distribution across users. To maintain consistency with our layer-based modeling, where user $i$ is served by a satellite in layer $l$, the JFI is evaluated using the corresponding user-layer throughput $R_{i,l}$. The modified JFI expression becomes~\cite{trankatwar2024power}:
\begin{equation}\label{eqn:JFI}
\text{JFI} = 
\frac{\left( \sum_{\ell \in \mathcal{K}} \sum_{i \in \mathcal{U}_\ell} R_{i,\ell} \right)^2}
{n \cdot \sum_{\ell \in \mathcal{K}} \sum_{i \in \mathcal{U}_\ell} R_{i,\ell}^2},
\end{equation}
\noindent
where $n$ is the total number of users across all layers.
The theoretical range of JFI is $\text{JFI} \in \left[ \frac{1}{n},\ 1 \right]$. $\text{JFI} = 1$ indicates perfect fairness — all users receive equal throughput. $\text{JFI} = \frac{1}{n}$ (minimum value) occurs when only one user gets all the resources and the others get none.

\subsection{Multi-Objective Optimization Formulation}
The problem aims to maximize two key performance objectives simultaneously: 
(i) the total system throughput, measured as the sum rate across all users, and 
(ii) fairness among users, quantified by JFI. 
Formally, the MOO problem is given by
\begin{subequations}
\begin{align}
(P1): & \max_{P_k, S_k} \; \big(R_{\text{total}}, \; \text{JFI}\big), \label{eqn:P1a}\\
\text{s.t.} \;\;  
& \sum_{j \in \mathcal{B}_s} \mathbf{1}\!\left[ \sum_{i \in \mathcal{U}} A_{ij} \geq 1 \right] \leq X, \quad && \forall s \in \mathcal{S}, \label{eqn:P1b} \\
& \sum_{i \in \mathcal{U}} \mathbf{1}_{[A_{ij} = 1]} \leq Z, \quad && \forall j \in \mathcal{B}_s , \label{eqn:P1c}\\
& A_{ij} \in \{0,1\}, \quad && \forall i,j, \label{eqn:P1d}
\end{align}    
\end{subequations}
where \(R_{\text{total}}\) and JFI denote the throughput and fairness objectives, respectively. 
Constraint~(\ref{eqn:P1b}) limits the number of active beams per satellite to at most \(X\). 
Constraint~(\ref{eqn:P1c}) restricts the number of users per beam to at most \(Z\), ensuring load balancing. 
Constraint~(\ref{eqn:P1d}) enforces binary user--beam assignments.

\noindent
The conflicting objectives of maximizing total throughput ($R_{\text{total}}$) and ensuring fairness (JFI) are unified into a single scalar objective using the \textit{weighted sum method}~\cite{trankatwar2024power}:
\begin{equation}\label{eqn:WSM}
f_i = \omega R_{\text{total}} + (1 - \omega)\,\text{JFI},
\end{equation}
where $\omega \in [0,1]$ governs the trade-off between throughput and fairness. A higher $\omega$ emphasizes throughput, while a lower value favors fairness. In this study, $\omega$ is fixed according to design preference, ensuring interpretability and aligning the optimization with deployment-specific quality-of-service goals.

\subsection{Assumptions}
\noindent
To maintain computational efficiency and realism, the following assumptions are adopted in multi-Layer NTN Modeling:
\begin{itemize}
    \item Constellation Model: Satellites follow \textit{Walker-Delta} configurations; each layer (LEO, MEO, GEO) is defined by its orbital planes and satellites per plane, uniformly distributed across orbits.
    \item User Mobility: Users are generated via a \textit{parametric urban--rural mobility model} within a fixed geographic region.
    \item Channel Model: Path loss includes \textit{FSPL} and \textit{EPL}; interference accounts for \textit{intra- and inter-satellite} effects; thermal noise follows \textit{Boltzmann’s formulation} with fixed receiver temperature and noise figure.
\end{itemize}
These assumptions balance \textit{accuracy and scalability}, consistent with prior NTN simulation studies.

\section{Solution for the MOO Problem}\label{Sol}
Following the MOO problem formulation, the proposed FTA-NTN algorithm~\ref{alg:algorithm1} jointly optimizes system throughput and fairness by simulating a realistic multi-layer NTN and employing \textit{Bayesian optimization} to identify optimal constellation configurations. The algorithm operates in two stages: (i) performance evaluation of a candidate configuration (Section~\ref{PartA}) and (ii) an outer-loop \textit{Bayesian optimization} process that searches the design space for the optimal constellation (Section~\ref{PartB}). 
\begin{algorithm}[htbp]
\caption{FTA-NTN Constellation Optimization}
\label{alg:algorithm1}
\KwIn{NTN parameters, $\mathcal{U}$, $X$, $Z$, $\omega$}
\KwOut{Optimal parameters $\mathcal{L}^* = (P_k^*, S_k^*)$}

\BlankLine
Define search space $\mathcal{L} = (P_k, S_k)$\;

\For{each trial $i$}{
    Sample $\mathcal{L}^{(i)} \in \mathcal{L}$\;
    Initialize Initialize covered user set $\mathcal{C} \gets \emptyset$\;
    
    \For{each time step $t$}{
        Get user and satellite positions - $\mathbf{x}_i(t)$, $\mathbf{x}_s$\;

        \For{each layer $k \in \{L, M, G\}$}{
            $\mathcal{U}_{\text{rem}}^{(t)} \gets \mathcal{U} \setminus \mathcal{C}$\;
            Determine feasible number of beams $K^{(k)}$\;
            Cluster $\mathcal{U}_{\text{rem}}^{(t)}$ into $K^{(k)}$ beams via KMeans\;

            \For{each cluster $\mathcal{C}_b$}{
                Find $s^*$ with remaining beam slots\;
                \If{valid $s^*$ exists}{
                    Assign $\mathcal{C}_b$ to $s^*$\;
                    Compute SINR for users in $\mathcal{C}_b$\;
                    Select top $Z$ users and update $\mathcal{C}$\;
                }
            }
        }
    }
    Compute $f_i = w \cdot R_{\text{total}} + (1 - w) \cdot \text{JFI}$ (Eq. \eqref{eqn:WSM})\;
}
\Return{$\mathcal{L}^* = \arg\max f_i$}
\end{algorithm}
\subsection{Performance Evaluation per Constellation}\label{PartA}
At each simulation time step $t$, the algorithm~\ref{alg:algorithm1} performs the following:

\subsubsection{Satellite Propagation and User Mobility}
Satellite positions are updated from the Walker Delta constellation model, and user locations are extracted from the parametric mobility model constrained to the Canadian landmass. 

\subsubsection{Beam Clustering and Assignment Strategy}
At each time step, FTA-NTN executes a three-stage strategy for beam clustering and user assignment.

\textbf{Step 1: Clustering Uncovered Users-}
At time step $t$, let $\mathcal{U}_{\text{rem}}^{(t)} \subseteq \mathcal{U}$ represent the set of users not yet served. For each satellite layer $k$, we denote the set of available satellites as $\mathcal{S}_k$, where each satellite can form at most $X^{(k)}$ beams, and each beam can serve up to $Z$ users.
To satisfy (\ref{eqn:P1b}) and (\ref{eqn:P1c}) constraints, we determine the feasible number of clusters as
\begin{align}
K_{\text{max}}^{(k)} &= |\mathcal{S}_k| \cdot X^{(k)} \\
K_{\text{min}}^{(k)} &= \left\lceil \frac{|\mathcal{U}_{\text{rem}}^{(t)}|}{Z} \right\rceil \\
K^{(k)} &= \min \left( K_{\text{max}}^{(k)},\ K_{\text{min}}^{(k)},\ |\mathcal{U}_{\text{rem}}^{(t)}| \right)
\end{align}

We apply K-Means clustering to the geospatial coordinates of users in $\mathcal{U}_{\text{rem}}^{(t)}$, resulting in $K^{(k)}$ clusters $\mathcal{C}_1, \dots, \mathcal{C}_{K^{(k)}}$, where each cluster $\mathcal{C}_b \subseteq \mathcal{U}_{\text{rem}}^{(t)}$ corresponds to a candidate beam footprint for subsequent assignment.

\textbf{Step 2: Beam-to-Satellite Assignment-}
For each cluster $\mathcal{C}_b$, the geographic centroid $\mathbf{x}_b$ is mapped to the nearest satellite $s \in \mathcal{S}_k$ that still has available beam capacity. The assignment is determined by minimizing the haversine distance:
\begin{equation}
s^* = \arg\min_{\substack{s \in \mathcal{S}_k \\ B_s^{(t)} < X^{(k)}}} d_{\text{km}}(\mathbf{x}_b, \mathbf{x}_s)
\end{equation}
where $B_s^{(t)}$ denote the number of beams currently utilized by satellite $s$ at time step $t$, $\mathbf{x}_b$ and $\mathbf{x}_s$ are the centroid and satellite coordinates, respectively. This procedure ensures that clusters are mapped to the nearest feasible satellite while respecting the maximum beams-per-satellite constraint.

\textbf{Step 3: SINR-Based User Selection-}
For each cluster $\mathcal{C}_b$, we compute the SINR of every user and select the top $Z$ users with the highest SINR values. These users are considered covered and are added to the served user set for layer $k$:
\begin{equation}
\mathcal{U}_k^{(t)} \gets \arg\max_{\substack{u \in \mathcal{C}_b}} \text{SINR}_u \quad \text{subject to} \quad |\mathcal{U}_k^{(t)}| \leq Z
\end{equation}
The covered users are removed from $\mathcal{U}_{\text{rem}}^{(t)}$, and beam usage counters for satellites are updated.

\subsubsection{SINR, Throughput, and Fairness Computation}
For each user within a beam footprint, SINR, throughput, and fairness are evaluated as described in Section~\ref{PL}, then aggregated over time to yield overall system performance in total throughput and fairness.

\subsection{Bayesian Optimization for Constellation Search}
\label{PartB}
The objective function $f_i$ (Eq.~\ref{eqn:WSM}) serves as the surrogate function for \textit{Bayesian optimization}.
To solve the scalarized problem, we employ \textit{Bayesian optimization}~\cite{snoek2012practical} in the outer loop of Algorithm~\ref{alg:algorithm1}. The search space is defined over constellation design parameters, namely $P_k$ and $S_k$ in the LEO and MEO layers. In each trial, a candidate configuration $\mathcal{L}_i$ is sampled, evaluated through system-level simulation with dynamic user mobility and beam allocation, and its objective value $f_i$ is recorded. Iterations update the surrogate model to balance exploration and exploitation of the design space. After sufficient trials, the optimal configuration is identified as $\mathcal{L}^* = \arg\max f_i,$ representing the constellation that jointly maximizes throughput and fairness under realistic system constraints.

\section{Numerical Results}\label{Res}
\noindent
This section presents the simulation results of the proposed FTA-NTN. 
The simulation parameters in Table~\ref{tab:sim_params} follow 3GPP NTN specifications~\cite{3gpp38811,3gpp38821} and are consistent with representative link-budget assumptions in recent NTN studies~\cite{9149179}. 
We adopt the 3GPP calibration scenario with an S-band carrier at 2.2~GHz and 20~MHz bandwidth, as defined in ~\cite{3gpp38811} for NTN system evaluation. 
The orbital altitude and inclination values are taken from ~\cite{3gpp38811,3gpp38821}, ensuring that the simulated LEO, MEO, and GEO layers reflect realistic deployment settings. 
Moreover, the link-budget analysis in~\cite{9149179} validates the realism of comparable assumptions in Ka-band, confirming the credibility of our setup across frequency ranges. 
A total of $N=500$ users are simulated to capture dense regional demand over 24 hours at hourly resolution. Each satellite employs $X=15$ beams with up to $Z=20$ users per beam, as defined in $(P1)$. Together, these parameters ensure that the evaluated FTA-NTN configurations remain standards-compliant and representative of practical NTN deployments.

\begin{figure*}[t]
  \centering
  {\includegraphics[width=0.31\textwidth]{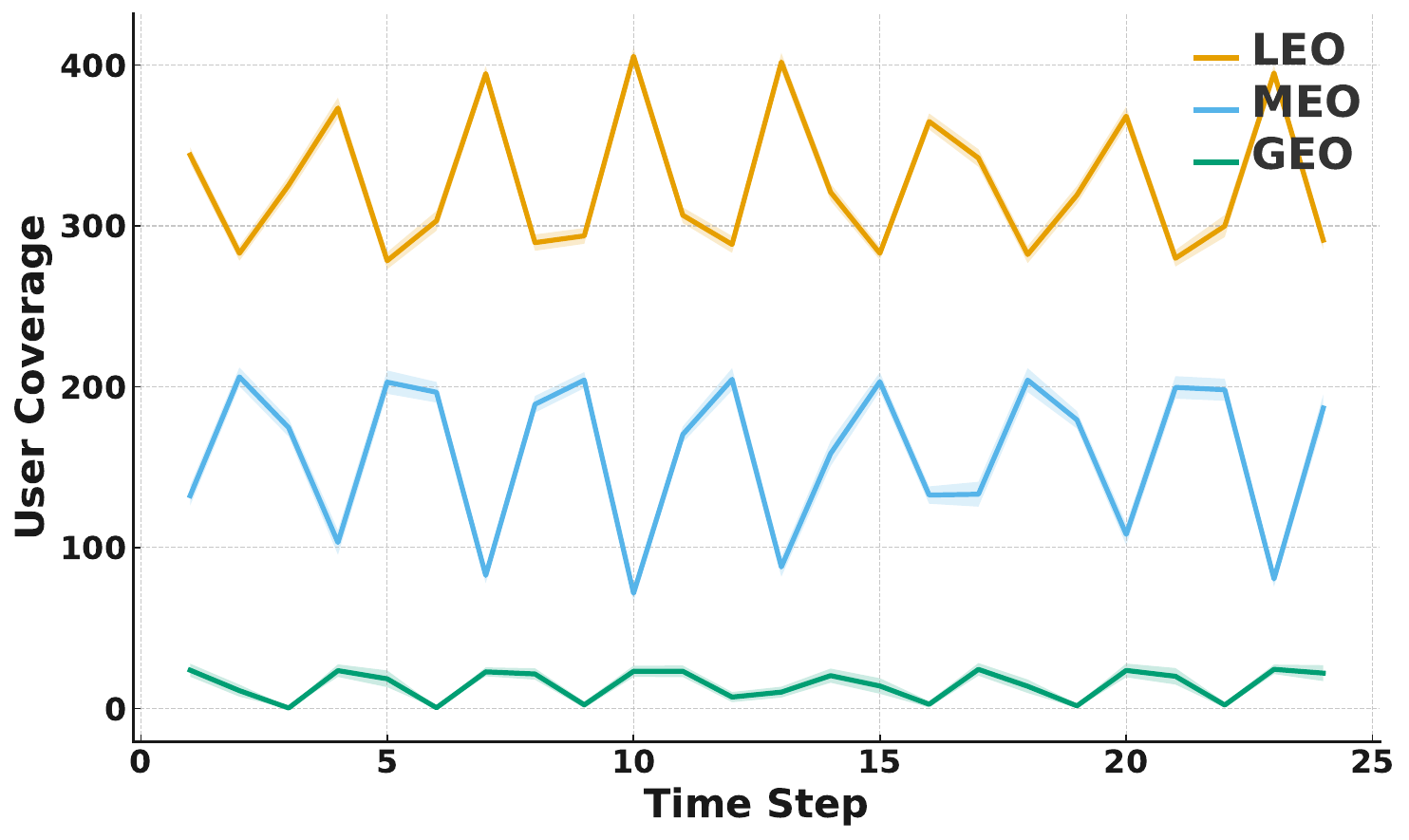}}\hfill
  {\includegraphics[width=0.31\textwidth]{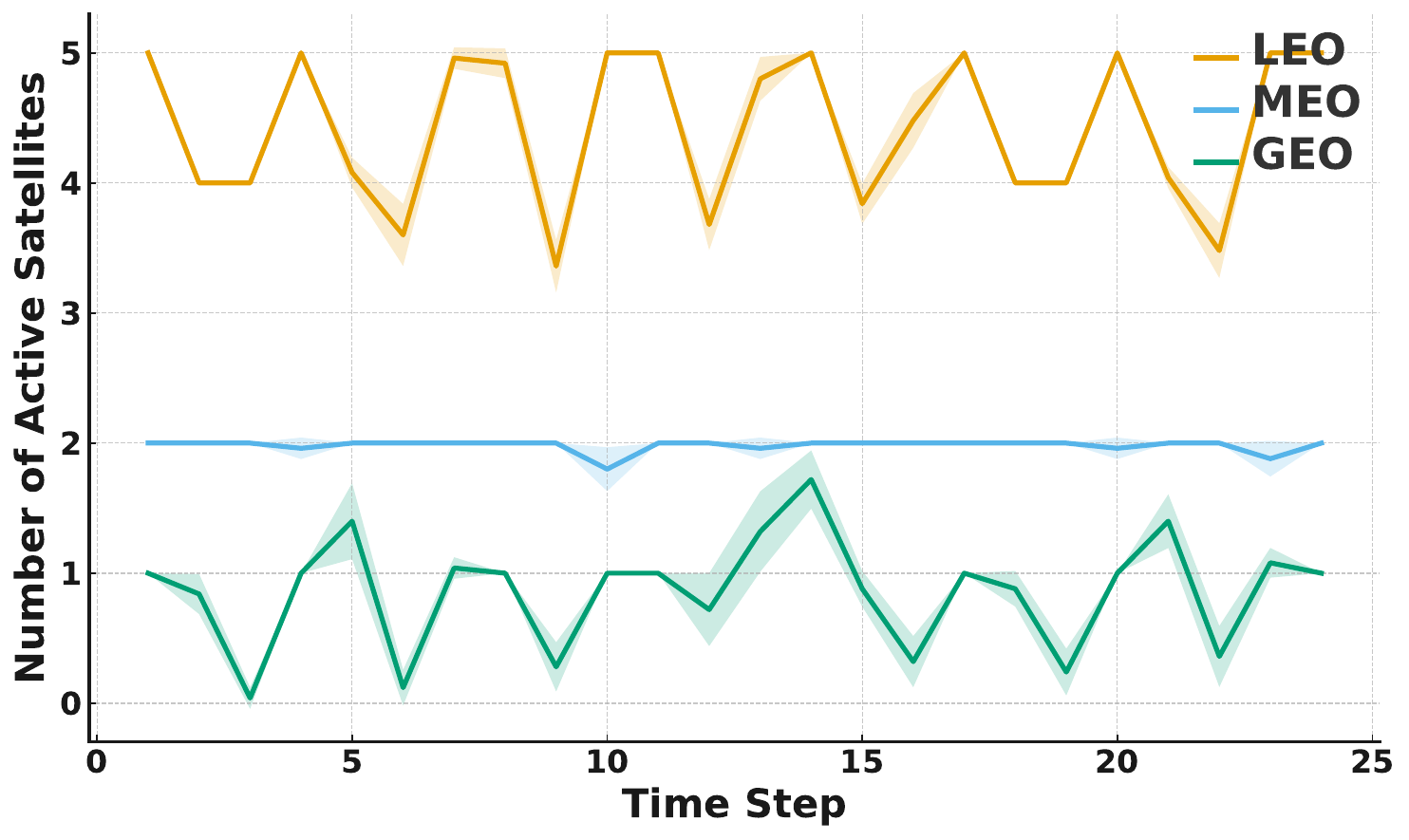}}\hfill
 {\includegraphics[width=0.31\textwidth]{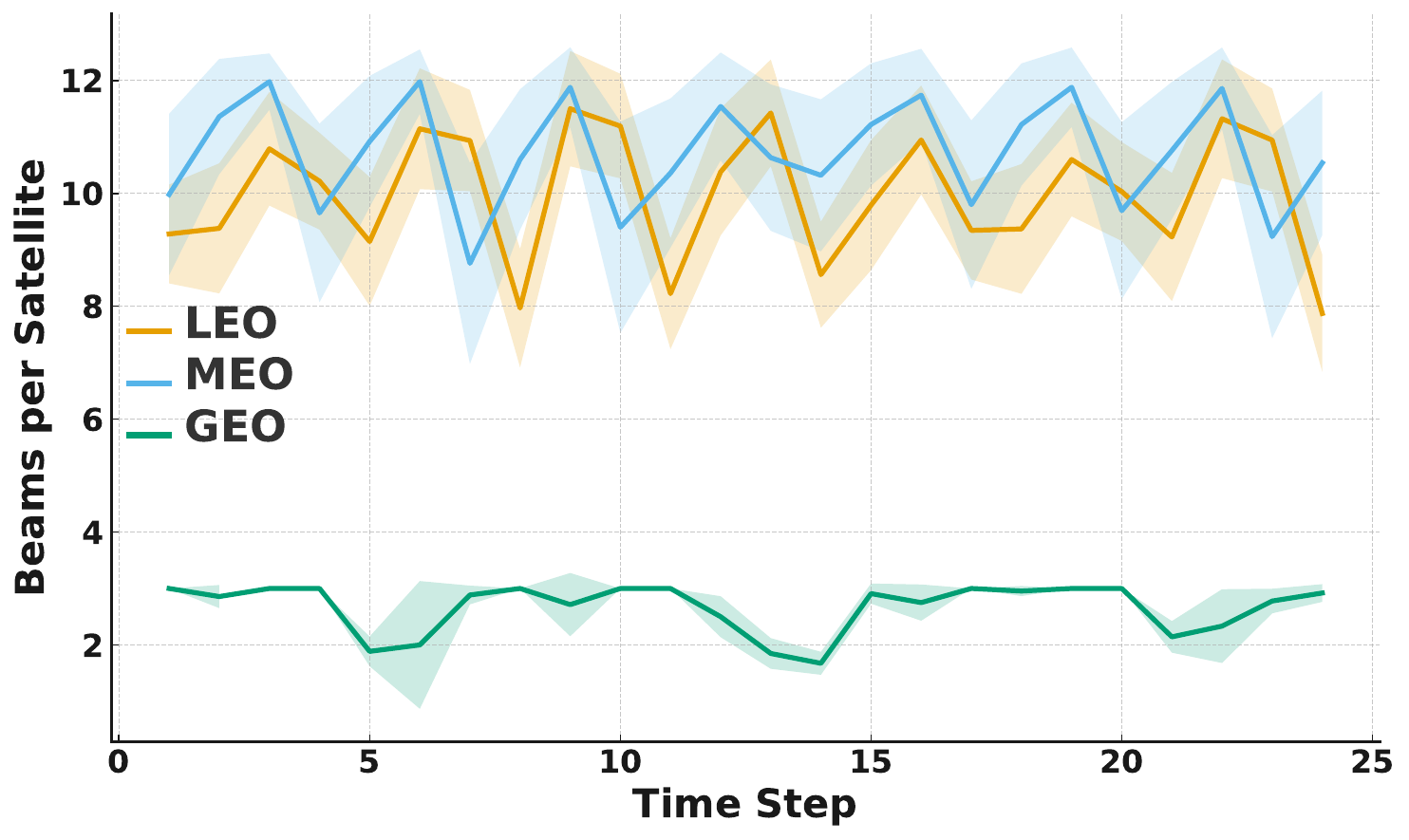}}
  \caption{Performance metrics across LEO, MEO, and GEO layers over time: (a) number of users covered, (b) active satellites per layer, and (c) average beams utilized per satellite. }
  \label{fig:result}
\end{figure*}

\noindent
The NTN constellation is initialized with a search space of $2$--$10$ orbital planes and $2$--$15$ satellites per plane for both LEO and MEO. The proposed FTA-NTN algorithm converges to an optimal configuration of $P_L=9$, $S_L=15$ for LEO and $P_M=7$, $S_M=3$ for MEO, achieving an average sum-rate of $9.88~\text{Gbps}$ and an average fairness of $0.42$ per time step. \noindent
Network performance is evaluated over $50$ realizations to ensure statistical robustness. Fig.~\ref{fig:result} presents performance metrics across LEO, MEO, and GEO: (a) users covered, (b) active satellites per layer, and (c) beams utilized per satellite. Solid curves denote mean values, while shaded regions indicate $95\%$ confidence intervals (CI), capturing variability across realizations.
\begin{table}[H]
\centering
\caption{Simulation and constellation parameters.}
\label{tab:sim_params}

\begin{tabular}{ll}
\hline
\multicolumn{2}{c}{\textbf{Global \& Radio Parameters}} \\
\hline
Earth radius $(R_{\oplus})$   & 6371 km \\
Users simulated $(N)$         & 500 \\
Simulated duration            & 24 hours \\
Number of time steps          & 24 \\
Number of realizations        & 50 \\
Weighting coefficient $(\omega)$ & 0.5 \\
Carrier frequency $(f_{\text{GHz}})$ & 2.2 GHz \\
Channel bandwidth $(B)$       & 20 MHz \\
Transmit power $(P_t)$        & 40 dBm \\
Satellite antenna gain $(G_t)$& 30 dBi \\
User received gain $(G_r)$    & 0 dBi \\
Receiver noise figure $(NF)$  & 2 dB \\
Beams per satellite $(X)$     & 15 \\
Users per beam $(Z)$      & 20 \\
\hline
\end{tabular}

\begin{tabular}{lccc}
\hline
\multicolumn{4}{c}{\textbf{Constellation Parameters}} \\
\hline
Layer & Altitude (km) & Inclination ($^\circ$) & Search Space / Configuration \\
\hline
LEO & 600   & 53 & Planes: 2--10, Sats/plane: 2--15 \\
MEO & 20200 & 56 & Planes: 2--10, Sats/plane: 2--15 \\
GEO & 35786 & 0  & Fixed (1 plane, 3 satellites) \\
\hline
\end{tabular}
\end{table}

\noindent Fig.~2(a) illustrates user coverage across the LEO, MEO, and GEO layers over a $24$-hour simulation. The LEO layer dominates with $280$--$400$ users due to its dense constellation and frequent handovers. The MEO layer provides stable coverage for $100$--$210$ users through wider beams and slower motion, while the GEO layer serves fewer than $30$ users owing to its broad, sparsely overlapping footprints. Overall, LEO ensures extensive reach, MEO adds stability, and GEO offers persistent baseline coverage---highlighting their complementary roles in the FTA-NTN architecture. 

\noindent Fig.~2(b) depicts the number of active satellites across the LEO, MEO, and GEO layers over $24$~hours. The LEO layer exhibits the highest dynamics, fluctuating between $2$ and $5$ active satellites due to rapid motion and frequent handovers. The MEO layer remains steady with about $2$ active satellites, while GEO activity is minimal ($0$--$1$) owing to its fixed position and wide footprints. Collectively, LEO offers agility and capacity, MEO provides stable mid-altitude coverage, and GEO ensures persistent but limited support, reflecting a balanced multi-layer NTN design.

\noindent Fig.~2(c) shows the average beams utilized per satellite over $24$~hours for FTA-NTN. The LEO layer dynamically employs $9$--$11$ beams due to frequent handovers, while the MEO layer operates within $9$--$12$ beams owing to wider footprints and fewer satellites. The GEO layer remains steady at about $3$ beams, occasionally dropping to $2$. Beam utilization across all layers stays within the capacity limit ($X=15$), validating Constraint~(\ref{eqn:P1b}). The resulting average beam radii---$96.4~\text{km}$ (LEO), $457.9~\text{km}$ (MEO), and $1151.2~\text{km}$ (GEO)---align with 3GPP NTN standards, confirming the physical consistency of the optimized model.
\begin{figure}[htbp]
    \centering
    \includegraphics[width=0.75\linewidth]{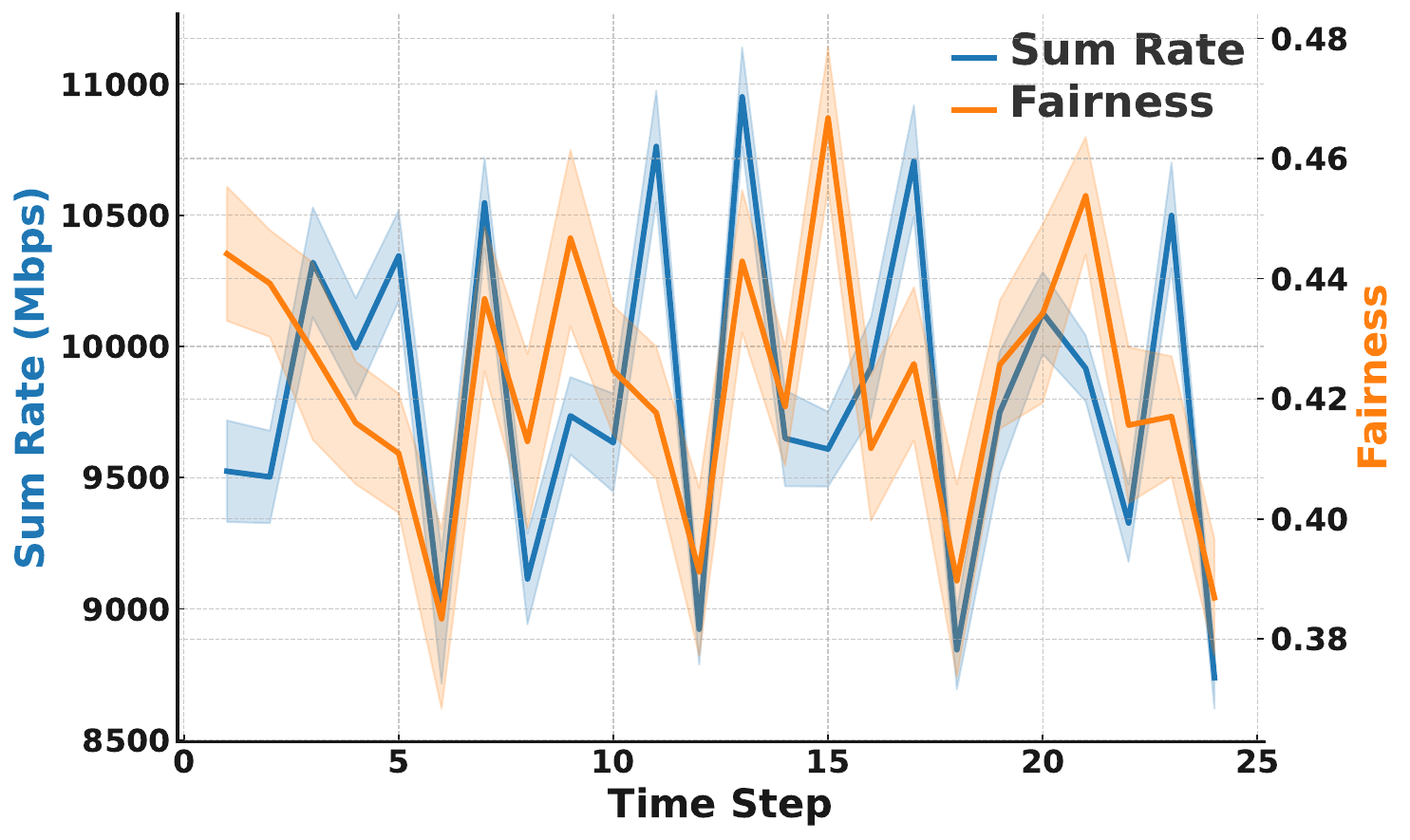}
    \caption{Sum rate and fairness over time.}
    \label{fig:SRandFairness}
\end{figure}
\noindent
Fig.~\ref{fig:SRandFairness} presents the temporal evolution of the system sum rate and JFI over 24 hours. The sum rate fluctuates between 8.5~Gbps and 11~Gbps, reflecting variations in satellite visibility and dynamic user demand. Fairness remains relatively stable around 0.4--0.46, indicating consistent resource distribution. Overall, FTA-NTN maintains a balanced performance, sustaining high capacity without compromising fairness.
\vspace{-0.4em}
\section{Conclusions}\label{Conc}
This paper introduced FTA-NTN, a MOO framework for NTNs that jointly maximizes throughput and fairness under realistic system constraints. By integrating multi-layer Walker Delta constellations, mobility-aware user modeling, adaptive K-Means beamforming, and Bayesian optimization, FTA-NTN effectively explores the throughput–fairness trade-off. Simulations with $500$ users achieved an aggregate throughput of $9.88~\text{Gbps}$ and an average fairness of $0.42$, converging to an optimal configuration of $9$ planes and $15$ satellites per plane in LEO, and $7$ planes with $3$ satellites per plane in MEO. These results, consistent with 3GPP NTN references, confirm FTA-NTN’s scalability and practicality for next-generation deployments. Future work will extend the framework to incorporate HAPs for enhanced multi-layer NTN flexibility.

\section*{Acknowledgment}
This work is supported in part by MITACS Accelerate Program project IT43178 and in part by the Natural Sciences and Engineering Research Council (NSERC) DISCOVERY and CREATE TRAVERSAL programs.
\bibliographystyle{ieeetr}

\end{document}